\documentclass[journal,a4paper]{IEEEtran}
\usepackage[T1]{fontenc}
\usepackage{amsmath,amssymb,graphicx,booktabs,array}
\usepackage{cite}
\usepackage[table]{xcolor}
\usepackage[hidelinks]{hyperref}
\usepackage{stfloats}
\usepackage{placeins}
\graphicspath{{figures/}}
\newcommand{\OPFCap}{65.572}
\newcommand{\HOPFCap}{56.892}
\newcommand{\CapReduction}{13.24}
\newcommand{\OPFIHD}{4.020}
\newcommand{\OPFTHD}{4.713}
\newcommand{\OPFVmin}{0.940}
\newcommand{\OPFVmax}{1.060}
\newcommand{\HOPFIHD}{2.789}
\newcommand{\HOPFTHD}{3.360}
\newcommand{\HOPFVmin}{0.978}
\newcommand{\HOPFVmax}{1.060}
\newcommand{\AHFIHD}{2.625}
\newcommand{\AHFTHD}{3.674}
\newcommand{\AHFVmin}{0.940}
\newcommand{\AHFVmax}{1.060}

\newcommand{\cG}{\mathcal G}
\newcommand{\cS}{\mathcal S}
\newcommand{\cH}{\mathcal H}
\newcommand{\cT}{\mathcal T}

\newcommand{\jmathu}{\mathrm{j}}
\begin{document}
\title{Harmonic-Constrained Hosting Capacity of Active Distribution Networks Using ESS Compensation and Certified Adaptive Decision-Tree Envelopes}
\author{Pengyu~Ren,~\IEEEmembership{Student Member,~IEEE,}
Wei~Sun\textsuperscript{*},~\IEEEmembership{Member,~IEEE,}
Fei~Teng,~\IEEEmembership{Senior Member,~IEEE,}
and~Gareth~Harrison,~\IEEEmembership{Senior Member,~IEEE}%
\thanks{Pengyu Ren, Wei Sun, and Gareth Harrison are with the School of Engineering, University of Edinburgh, Edinburgh, U.K. Fei Teng is with the Department of Electrical and Electronic Engineering, Imperial College London, London, U.K.\par\noindent\textsuperscript{*}Corresponding author: Wei Sun (e-mail: W.Sun@ed.ac.uk).}}
\maketitle
\begin{abstract}
Active distribution networks increasingly rely on coordinated inverter-based resources to accommodate renewable generation, while harmonic limits constrain the capacity that can be connected. This paper proposes a harmonic optimal power flow (HOPF) method with active energy-storage-system (ESS) support. ESS charging, discharging and harmonic-current compensation are jointly scheduled under state-of-charge dynamics and a shared converter-current rating. An adaptive decision tree learns voltage--current partitions from OPF operating points. Within each leaf, an affine plane is fitted on an analytic grid and certified as a lower envelope. A disaggregated mixed-integer formulation embeds these envelopes in a second-order-cone-relaxed network model. Annual hourly data are assessed through daily capacity optimization, followed by componentwise capacity aggregation and chronological full-year feasibility validation. In the case study, the OPF reference of \OPFCap{} MW exceeds the adopted harmonic bounds, whereas HOPF without harmonic support yields an annually validated capacity of \HOPFCap{} MW. With each generator capacity bounded by its OPF reference, the proposed method recovers the complete reference allocation. Coordinated energy dispatch and harmonic compensation thus recover generation capacity within the installed ESS converter ratings.

\end{abstract}

\section*{Nomenclature}
\begingroup
\small
\newenvironment{symbolentries}{%
  \begin{list}{}{%
    \setlength{\labelwidth}{25mm}%
    \setlength{\labelsep}{2mm}%
    \setlength{\leftmargin}{27mm}%
    \setlength{\rightmargin}{0pt}%
    \setlength{\itemsep}{1pt}%
    \setlength{\parsep}{0pt}%
    \setlength{\topsep}{3pt}%
    \renewcommand{\makelabel}[1]{##1\hfil}}%
  \raggedright
}{\end{list}}

\subsection*{Abbreviations}
\begin{symbolentries}
\item[ADN] Active distribution network.
\item[DT] Decision tree.
\item[ESS] Energy storage system.
\item[HC] Hosting capacity.
\item[IBR] Inverter-based resource.
\item[IHD, THD] Individual and total harmonic distortion of voltage.
\item[OLTC] On-load tap changer.
\item[OPF, HOPF] Optimal power flow; harmonic OPF.
\item[SOC] State of charge: $100E_{st}/\bar E_s$.
\item[SOCP, MISOCP] Second-order cone programming; mixed-integer SOCP.
\end{symbolentries}

\subsection*{Sets and Indices}
\begin{symbolentries}
\item[$\mathcal B,\mathcal L$] Buses and directed branches; $i,j$ index buses and $\ell=(i,j)$ a branch.
\item[$\cG,\cS$] IBR generators and ESS units, indexed by $g$ and $s$.
\item[$\cG_i,\cS_i$] Generators and ESS units connected at bus $i$.
\item[$\cG_{\!B}$] Buses containing modeled IBR generators.
\item[$\cT,\cH$] Operating hours and harmonic orders, indexed by $t$ and $h$.
\item[$\mathcal D,d$] Set of operating days and day index.
\item[$\delta^+(i),\delta^-(i)$] Outgoing and incoming branches at bus $i$.
\item[$\mathcal M_\ell,m$] Available OLTC tap positions and their index.
\item[$r,\mathcal R_r$] DT leaf index and its voltage--current rectangle.
\end{symbolentries}

\subsection*{Network and Resource Parameters}
\begin{symbolentries}
\item[$S_{\mathrm b},\Delta t$] Base power (MVA) and time-step duration (h).
\item[$a_{gt},\epsilon$] Available output per unit of installed capacity and daily curtailment fraction limit.
\item[$r_\ell,x_\ell,a_\ell$] Fundamental series resistance, reactance, and transformer tap ratio.
\item[$b_i^{\mathrm{sh}},\bar S_\ell$] Bus-side charging susceptance and branch apparent-power limit.
\item[$\underline V,\overline V$] Fundamental voltage-magnitude limits.
\item[$p^d_{it},q^d_{it}$] Active and reactive load demand.
\item[$\bar p_s,\bar E_s,\bar I_s$] ESS charge/discharge power limit, energy capacity (MWh), and converter RMS-current rating.
\item[$\eta_{\mathrm{ch}},\eta_{\mathrm{dis}}$] Charging and discharging efficiencies.
\item[$\alpha_h,h_c$] Harmonic-to-fundamental emission ratio and compensated order ($h_c=5$).
\item[$L_h,L_{\mathrm{THD}}$] IHD and THD limits as voltage ratios, not percentages.
\item[$\gamma$] Numerical safety factor on harmonic limits.
\item[$R^h_\ell,X^h_\ell$] Branch resistance and reactance evaluated at harmonic order $h$.
\item[$Y^h_t,Z^h_t$] Harmonic bus admittance and impedance matrices; $Z^h_t=(Y^h_t)^{-1}$.
\item[$Z^1_{it},Y^{h,\mathrm{load}}_{it}$] Nominal-voltage fundamental load impedance and harmonic load admittance.
\end{symbolentries}

\subsection*{Planning and Operating Quantities}
\begin{symbolentries}
\item[$C_g$] Installed generator capacity, shared by the hours of a daily subproblem; $S_{\mathrm b}C_g$ is in MW.
\item[$C_{g,d}^{\star},C_g^{\mathrm{ann}}$] Daily optimal capacity and annually validated common capacity of generator $g$.
\item[$C_g^{\mathrm{OPF}}$] OPF reference capacity of generator $g$.
\item[$\tau_{\ell t},\tau_{\ell m}$] Operating reciprocal tap ratio and candidate ratio at tap position $m$.
\item[$y_{\ell mt},w_{\ell mt}$] Binary tap selector and product $v_{it}y_{\ell mt}$.
\item[$p_{gt},c_{gt}$] Actual generator active power and curtailed available power.
\item[$P_{jt}$] Total IBR active output at bus $j$.
\item[$v_{it},\widetilde v_{\ell t}$] Squared fundamental bus voltage and tap-adjusted sending voltage $v_{it}/a_\ell^2$.
\item[$v_{st}$] Squared fundamental voltage at the bus hosting ESS unit $s$.
\item[$\ell_{\ell t}$] Squared branch series-current magnitude.
\item[$p^f_{\ell t},q^f_{\ell t}$] Active/reactive series power entering the branch at the sending end.
\item[$p^r_{\ell t},q^r_{\ell t}$] Active/reactive series power entering the branch at the receiving end.
\item[$p^{\mathrm{ext}}_{it},q^{\mathrm{ext}}_{it}$] Active/reactive injection from the supply or external interconnector.
\item[$p^{\mathrm{ch}}_{st},p^{\mathrm{dis}}_{st}$] ESS charging and discharging powers.
\item[$p^e_{st},u_{st}$] Net ESS discharge power and mode binary ($1$: charging; $0$: discharging).
\item[$E_{st}$] Stored energy at the end of hour $t$ (MWh); $E_{s0}$ is initial energy.
\item[$k_{st}$] ESS compensation-current magnitude at the selected harmonic order.
\item[$I^0_{jt},I^{0,\mathrm{ex}}_{jt}$] Conservative fundamental source current and recalculated value $P_{jt}/\sqrt{v_{jt}}$.
\item[$J^h_{jt}$] Residual harmonic source-current magnitude after local compensation.
\item[$\bar V^h_{it},\bar V^{h,\mathrm{ex}}_{it}$] Harmonic voltage-magnitude bounds using conservative and recalculated source currents.
\item[$\mathrm{IHD}^h_{it},\mathrm{THD}_{it}$] Individual and RMS-aggregated harmonic voltage ratios to fundamental voltage, reported in percent.
\end{symbolentries}

\subsection*{Adaptive DT Envelope}
\begin{symbolentries}
\item[$f(v,I),\bar I^0$] Analytic surface $\sqrt v I$ and upper current bound of the envelope domain.
\item[$v_r^-,v_r^+,I_r^-,I_r^+$] Voltage-squared and current bounds of leaf $r$.
\item[$a_r,b_r,c_r,\widehat c_r$] Affine coefficients, fitted intercept, and certified intercept.
\item[$\rho_r,\varepsilon_c$] Maximum fitted residual over a leaf and downward certification margin.
\item[$q_r,L_{r1},L_{r2}$] Square-root chord and fallback affine lower bounds in leaf $r$.
\item[$\underline f_r(v,I)$] Certified affine lower bound on $\sqrt v I$ in leaf $r$.
\item[$z_{jtr}$] Binary selector for leaf $r$ at bus $j$, hour $t$.
\item[$v_{jtr},I_{jtr}$] Leaf-disaggregated squared voltage and fundamental current.
\end{symbolentries}
\endgroup

\section{Introduction}

\IEEEPARstart{A}{ctive} distribution networks (ADNs) accommodate renewable generation through coordinated operation of distributed resources, linking network hosting capacity to the operational flexibility available after connection~\cite{ochoa2010minimizing}. Optimal power flow (OPF) provides a framework for assessing this capacity under network and device constraints. However, capacity assessments based only on fundamental-frequency voltage and thermal limits can overlook harmonic restrictions associated with inverter-based resources (IBRs)~\cite{sun2022distribution}. With appropriate control, grid-connected inverters can also compensate harmonic currents alongside renewable-power injection~\cite{singh2010grid}. When energy-storage-system (ESS) converters provide this support, their compensation capability must be coordinated with storage dispatch and converter limits. The capacity benefit of ESS harmonic support therefore depends on both its dispatch and its available converter current.

Harmonic voltage distortion is a network-wide phenomenon arising from the interaction between harmonic-current injections and frequency-dependent network impedances~\cite{taghvaie2023comprehensive}. Currents emitted by power-electronic converters and nonlinear loads propagate through the network, with excessive distortion potentially causing equipment overheating and malfunction~\cite{singh2009power}. When multiple harmonic sources are present, the voltage distortion at a bus depends on their combined contributions and the network transfer impedances, rather than solely on the local injection. Consequently, the most exposed bus need not coincide with an IBR connection point. The planning implications have been examined by incorporating individual and total harmonic distortion limits into the selection of distributed generation (DG) locations and capacities~\cite{pandi2012determining}, highlighting the need to assess network-wide distortion when determining admissible generation capacity.

Impedance-scanning methods have been used to evaluate resonance participation in renewable energy systems~\cite{xue2023resonance}, while dynamic harmonic estimation has been combined with active-filter control for waveform distortion suppression~\cite{ornelas2019optimal}. These approaches support system characterization and device-level mitigation, whereas determining admissible generation requires explicit consideration of network operating constraints. At the network level, Anti\'c \emph{et al.}~\cite{antic2026envelopes} incorporate harmonics into a three-phase OPF model to calculate dynamic import and export operating envelopes. Such time-varying operating limits address a different decision from selecting an installed-capacity vector that remains fixed across the operating horizon. A capacity assessment must connect these hourly operating conditions to a common installed-capacity decision.

Harmonic-constrained capacity optimization couples several nonlinear relationships. The fundamental-frequency AC power-flow equations are nonconvex~\cite{low2014convex}. Although harmonic propagation is linear for fixed network impedances and prescribed current injections, the injections become coupled to operating decisions when their magnitudes depend on generation output and terminal voltage. When ESS operation is included, the formulation must also accommodate intertemporal energy constraints and discrete charging/discharging decisions. Existing solution methods either evaluate these relationships through external simulations or embed them directly in the optimization model.

Simulation-assisted optimization couples external search algorithms with power-flow evaluations. Misra \emph{et al.}~\cite{misra2017harmonic} combine a genetic algorithm in MATLAB with an unbalanced OpenDSS model to dispatch wind generation for harmonic reduction in feeders with electric-vehicle charging. More recently, Wang \emph{et al.}~\cite{wang2024pvess} employ accelerated particle swarm optimization for PV--ESS harmonic-mitigation cost minimization, with MATLAB--OpenDSS calculations used in their case studies. Complementary work by Yang and Hsu~\cite{yang2023opendss} develops injected-current passive-filter models and an iterative OpenDSS interface for power-flow analysis, rather than an outer capacity-optimization algorithm. Simulation-assisted methods accommodate detailed network models through repeated power-flow evaluations; their computational cost grows with the number of candidate decisions and operating conditions evaluated.

Direct formulations instead represent harmonic network and device relations explicitly within the optimization problem. Wang \emph{et al.}~\cite{wang2018optimal} develop a semidefinite programming (SDP) relaxation for converter-based harmonic and unbalance compensation, targeting network losses and power-quality indices. Xiao \emph{et al.}~\cite{xiao2022optimal} construct a multiphase ESS compensation model and use current-injection equations and approximations to obtain convex quadratically constrained quadratic programming (QCQP) subproblems, with outer iterations to improve solution accuracy. Their ESS active-power schedule is supplied by day-ahead dispatch rather than jointly optimized with compensation. In contrast, Geth and Van Acker~\cite{geth2022harmonic} retain a continuous nonlinear programming formulation, representing transformer excitation through multivariate nonlinear splines. These formulations show how harmonic compensation can be incorporated into network optimization. Joint capacity and ESS planning additionally requires the source-current approximation to preserve the direction of the harmonic limits while coupling compensation to energy dispatch.

Table~\ref{tab:position} summarizes the objectives and operational decisions covered by these studies.

\begin{table*}[!b]
\caption{Comparison of Harmonic Optimization Methods for Distribution Networks}
\label{tab:position}\centering\small
\setlength{\tabcolsep}{2pt}
\renewcommand{\arraystretch}{1.3}
\begin{tabular}{@{}>{\raggedright\arraybackslash}p{.15\textwidth}>{\raggedright\arraybackslash}p{.18\textwidth}>{\raggedright\arraybackslash}p{.195\textwidth}*{4}{>{\centering\arraybackslash}p{.105\textwidth}}@{}}
\toprule
Study & Objective & Nonlinear treatment & Multiple harmonic sources & Active IBR harmonic support & Shared capacity decision & Joint SOC dispatch \\
\midrule
Wang et al.\newline (2019)~\cite{wang2018optimal} & Reduce losses and improve power quality & SDP relaxation & $\checkmark$ & $\checkmark$ & -- & -- \\
Sun \& Harrison\newline (2022)~\cite{sun2022distribution} & Maximize installed generation capacity & NLP / MINLP (OA) & $\checkmark$ & -- & $\checkmark$ & -- \\
Geth \& Van Acker\newline (2022)~\cite{geth2022harmonic} & Reduce harmonic voltage & NLP; excitation splines & $\checkmark$ & --$^c$ & -- & -- \\
Xiao et al.\newline (2022)~\cite{xiao2022optimal} & Reduce losses and improve power quality & Iterative convex QCQP & $\checkmark$ & $\checkmark$ & -- & --$^a$ \\
Wang et al.\newline (2024)~\cite{wang2024pvess} & Minimize harmonic mitigation cost & NLP (APSO) & $\checkmark^b$ & $\checkmark$ & -- & --$^a$ \\
Anti\'c et al.\newline (2026)~\cite{antic2026envelopes} & Maximize export/import power & Three-phase NLP & $\checkmark$ & -- & -- & -- \\
\textbf{This work} & \textbf{Maximize installed generation capacity} & \textbf{Certified DT--MISOCP} & $\boldsymbol{\checkmark}$ & $\boldsymbol{\checkmark}$ & $\boldsymbol{\checkmark}$ & $\boldsymbol{\checkmark}$ \\
\bottomrule
\end{tabular}
\par\smallskip
\begin{minipage}{\textwidth}\footnotesize
$\checkmark$: included; --: not jointly optimized/modeled. NLP: nonlinear programming; MINLP: mixed-integer NLP; OA: outer approximation; APSO: accelerated particle swarm optimization.
$^a$ESS active power~\cite{xiao2022optimal} or operating modes~\cite{wang2024pvess} are prescribed externally. $^b$Includes the distributed-background-harmonic experiment.
$^c$Uses a dedicated active power filter, not harmonic support from an IBR with a generation/storage function.
\end{minipage}
\end{table*}

This paper develops HOPF with active ESS support to maximize installed generation capacity while jointly scheduling storage energy and local harmonic compensation. A certified decision-tree (DT) envelope supplies conservative source currents, and a shared converter-current constraint links the two ESS services.

\begin{figure*}[!t]
    \centering
    \includegraphics[width=\textwidth]{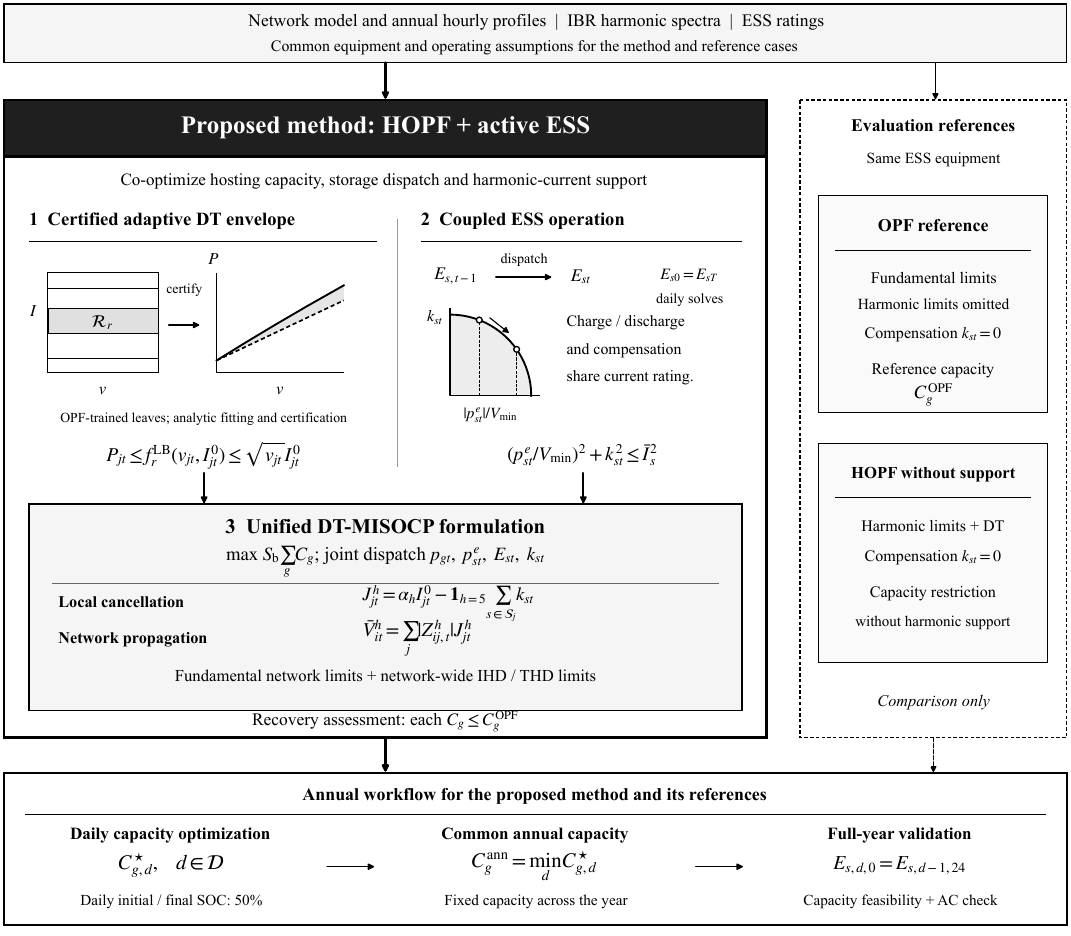}
    \caption{Proposed HOPF with active ESS support. OPF operating points define adaptive DT partitions; analytic leafwise fitting and certification provide conservative current envelopes. ESS energy scheduling and fifth-harmonic compensation share converter capability within the unified DT--MISOCP model. OPF and uncompensated HOPF provide evaluation references. Daily optimal capacities are aggregated componentwise and validated chronologically over the year with cross-day SOC continuity.}
    \label{framework}
\end{figure*}

\subsection{Contribution}
The main contributions are:

\begin{itemize}
    \item A unified HOPF method with active ESS support jointly optimizes generation capacity, storage dispatch and local fifth-harmonic compensation. SOC dynamics and the shared converter-current rating link the available harmonic service to energy scheduling.
    \item An adaptive DT trained on OPF operating points partitions the voltage--current domain. Analytic grid fitting and leafwise certification construct conservative affine envelopes, which are embedded through a disaggregated mixed-integer conic formulation.
    \item A full-year assessment combines daily capacity optimization with chronological validation of a common installed-capacity vector. The case study demonstrates recovery of the OPF reference allocation under per-generator capacity upper bounds, supported by network-wide harmonic indicators, ESS schedules and nonlinear AC power-flow validation.
\end{itemize}

\section{ADN Planning and Operational Model}\label{sec:network}
Fig.~\ref{framework} summarizes the proposed HOPF with active ESS support. The method has an offline envelope-construction stage and a capacity-assessment stage. Offline, fundamental OPF solutions under varying load and generation conditions supply voltage, current and power samples. A regression tree uses these samples to partition the voltage--current domain. Affine planes are then fitted to analytic samples within each leaf and shifted or replaced to obtain certified lower bounds on the power surface. Section~\ref{sec:dt} gives the construction and its mixed-integer representation.

The capacity model combines these envelopes with the fundamental network equations, harmonic propagation constraints and ESS operating limits. Generation capacity is shared across the hours of each daily subproblem, while generation output, ESS dispatch and OLTC settings adapt to hourly conditions. The ESS converter supplies both fundamental active current and fifth-harmonic compensation; its current rating couples the two services, and SOC dynamics link dispatch across time. The envelopes provide conservative IBR source currents for the harmonic constraints in Section~\ref{sec:harmonic}.

Two reference cases isolate the effect of harmonic restrictions and the benefit of compensation: fundamental OPF and HOPF with ESS harmonic support disabled. All cases use the same storage equipment and optimize their own OLTC settings. In the proposed method, $C_g^{\mathrm{OPF}}$ denotes generator $g$'s OPF reference capacity, and $C_g\le C_g^{\mathrm{OPF}}$ is imposed while total capacity is maximized. Reaching all these bounds recovers the reference allocation. Finally, the componentwise minima of daily optimal capacities form one annual installation vector. Chronological full-year validation then checks its operation with SOC carried between days, as described in Section~\ref{sec:annual}.

\subsection{Notation and common-capacity objective}
Let $\mathcal B$ and $\mathcal L$ denote the bus and directed branch sets, $\cG$ the IBR set, $\cS$ the ESS set, and $\cT$ the hourly operating horizon. The subsets $\cG_i$ and $\cS_i$ contain devices at bus $i$. For a branch $\ell=(i,j)$, define the outgoing and incoming sets as $\delta^+(i)$ and $\delta^-(i)$. The harmonic orders are collected in $\cH$. Lower- and upper-case voltage symbols are distinguished: $V^1_{it}$ is the fundamental voltage phasor and $v_{it}=|V^1_{it}|^2$ its squared magnitude, whereas $\bar V^h_{it}$ is a harmonic-voltage magnitude bound.

Power and current variables in the network equations are per unit on consistent bases. The power base is denoted $S_{\mathrm b}$; ESS energy $E_{st}$ is in MWh and $\Delta t$ is in hours. Within a daily capacity subproblem, generator $g$ has one installed capacity $C_g$ shared by its 24 operating hours. Day indices are suppressed in the operational equations below; annual aggregation and validation are described in Section~\ref{sec:annual}. With availability $a_{gt}$, actual output $p_{gt}$, and curtailed output $c_{gt}$,
\begin{align}
\max\quad & S_{\mathrm b}\sum_{g\in\cG}C_g,\label{eq:objective}\\
p_{gt}&=a_{gt}C_g-c_{gt},\quad 0\le c_{gt}\le a_{gt}C_g,\label{eq:generation}\\
\sum_t c_{gt}\Delta t&\le\epsilon\sum_t a_{gt}C_g\Delta t,\quad g\in\cG.\label{eq:curtail}
\end{align}
The curtailment budget is imposed separately on each generator over the daily window, with $\epsilon=0.05$. Generation and ESS operate at unity fundamental power factor in the case study, with ESS providing controllable harmonic-current support.

\subsection{Branch-flow SOCP model and transformer treatment}
For branch $\ell=(i,j)$, let $r_\ell,x_\ell$ be the series parameters, $a_\ell$ the ideal tap ratio, $\ell_{\ell t}$ the squared series-current magnitude, and $p^f_{\ell t},q^f_{\ell t}$ the sending-end series powers. The receiving-end powers $p^r_{\ell t},q^r_{\ell t}$ are positive into the branch from bus $j$. With $\widetilde v_{\ell t}=v_{it}/a_\ell^2$,
\begin{align}
p^r_{\ell t}
&= -p^f_{\ell t}+r_\ell\ell_{\ell t},
\label{eq:loss}\\
q^r_{\ell t}
&= -q^f_{\ell t}+x_\ell\ell_{\ell t},\\
v_{jt}
&= \widetilde v_{\ell t}
   -2\bigl(r_\ell p^f_{\ell t}
          +x_\ell q^f_{\ell t}\bigr)
\notag\\
&\quad +(r_\ell^2+x_\ell^2)\ell_{\ell t}.
\label{eq:drop}
\end{align}
\begin{equation}
(p^f_{\ell t})^2+(q^f_{\ell t})^2
\le \widetilde v_{\ell t}\ell_{\ell t}.
\label{eq:socp}
\end{equation}
The last inequality is a rotated second-order-cone constraint. Both branch ends satisfy $\|(p^f,q^f)\|_2\le\bar S_\ell$ and $\|(p^r,q^r)\|_2\le\bar S_\ell$. The implementation additionally bounds $\ell_{\ell t}$ by $\min(a_\ell^2,1)\bar S_\ell^2/\underline V^2$.

Let $p^{\mathrm{ext}}_{it},q^{\mathrm{ext}}_{it}$ denote exchange at the supply bus or external interconnector, subject to their specified exchange bounds, and let $b_i^{\mathrm{sh}}$ collect incident bus-side charging susceptances. With specified loads $p^d_{it},q^d_{it}$, nodal balances are
\begin{align}
\sum_{g\in\cG_i}p_{gt}+\sum_{s\in\cS_i}p^e_{st}
+p^{\mathrm{ext}}_{it}-p^d_{it}
&=\sum_{\ell\in\delta^+(i)}p^f_{\ell t}+\sum_{\ell\in\delta^-(i)}p^r_{\ell t},\label{eq:pbal}\\
q^{\mathrm{ext}}_{it}-q^d_{it}+b_i^{\mathrm{sh}}v_{it}
&=\sum_{\ell\in\delta^+(i)}q^f_{\ell t}+\sum_{\ell\in\delta^-(i)}q^r_{\ell t}.\label{eq:qbal}
\end{align}
Here $p^e_{st}=p^{\mathrm{dis}}_{st}-p^{\mathrm{ch}}_{st}$. Voltage limits are $\underline V^2\le v_{it}\le\overline V^2$, and the supply-bus voltage is fixed to unity.

The optimization uses a branch-flow SOCP relaxation~\cite{low2014convex}. Nonlinear AC power-flow validation confirms the feasibility of the resulting capacity configurations.

OLTC settings are optimized separately in each scenario. On branch $\ell$, tap index $m\in\mathcal M_\ell$ identifies an available reciprocal ratio $\tau_{\ell m}$. Binary $y_{\ell mt}$ selects one ratio at hour $t$:
\begin{align}
\sum_{m\in\mathcal M_\ell}y_{\ell mt}&=1,\quad y_{\ell mt}\in\{0,1\},\label{eq:tap_select}\\
\tau_{\ell t}&=\sum_{m\in\mathcal M_\ell}\tau_{\ell m}y_{\ell mt},\label{eq:tap_ratio}\\
\widetilde v_{\ell t}&=\tau_{\ell t}^{2}v_{it},\quad \ell=(i,j).\label{eq:tap_voltage}
\end{align}
Here $\tau_{\ell t}=1/a_{\ell t}$, and $a_\ell$ in the branch equations denotes its operating value at hour $t$. Because each candidate ratio is constant, auxiliary variables $w_{\ell mt}=v_{it}y_{\ell mt}$ give the equivalent expression $\widetilde v_{\ell t}=\sum_m\tau_{\ell m}^2w_{\ell mt}$. The binary products are enforced using the voltage bounds:
\begin{align}
\underline V^2 y_{\ell mt}&\le w_{\ell mt}\le\overline V^2 y_{\ell mt},\\
\underline V^2(1-y_{\ell mt})&\le v_{it}-w_{\ell mt}
\le\overline V^2(1-y_{\ell mt}).
\end{align}
Thus the selected tap enters the conic network model through linear constraints. Non-OLTC branches have $\tau_{\ell t}=1$.

\subsection{ESS energy and operating modes}\label{sec:ess}
Within each daily capacity subproblem, binary $u_{st}=1$ enables charging and $u_{st}=0$ enables discharging:
\begin{align}
0\le p^{\mathrm{ch}}_{st}&\le\bar p_s u_{st},\label{eq:mode}\\
0\le p^{\mathrm{dis}}_{st}&\le\bar p_s(1-u_{st}),\\
E_{st}&=E_{s,t-1}+S_{\mathrm b}\Delta t
\left(\eta_{\mathrm{ch}}p^{\mathrm{ch}}_{st}
-\frac{p^{\mathrm{dis}}_{st}}{\eta_{\mathrm{dis}}}\right),\label{eq:soc}\\
0\le E_{st}&\le\bar E_s,\quad E_{s0}=E_{s,24}=0.5\bar E_s.\label{eq:terminal}
\end{align}
Here $\bar p_s$ and $\bar E_s$ are the storage power and energy limits, and $\eta_{\mathrm{ch}},\eta_{\mathrm{dis}}$ are the charging and discharging efficiencies. Storage dispatch is optimized in all three scenarios, with initial and final SOC set to 50\% in each daily capacity subproblem.

\subsection{Annual capacity aggregation and validation}\label{sec:annual}
The assessment uses a full year of hourly operating data. Daily capacity subproblems retain the intertemporal ESS constraints within each 24-hour window. Let $\mathcal D$ be the set of days and $C_{g,d}^{\star}$ the optimal capacity of generator $g$ in the subproblem for day $d$. The annual common-capacity vector is constructed componentwise as
\begin{equation}
C_g^{\mathrm{ann}}=\min_{d\in\mathcal D} C_{g,d}^{\star},\qquad g\in\cG.\label{eq:annualcapacity}
\end{equation}
Different generators may attain their minima on different days. The resulting vector specifies one installation that is held fixed throughout chronological full-year validation. Operating decisions are determined for this common configuration under the hourly network, harmonic and ESS constraints. During validation, the final energy of each day supplies the initial energy of the next:
\begin{equation}
E_{s,d,0}=E_{s,d-1,24},\qquad d=2,\ldots,|\mathcal D|.\label{eq:soccarry}
\end{equation}
The resulting quantity $S_{\mathrm b}\sum_g C_g^{\mathrm{ann}}$ is reported as the annually validated capacity. The daily solutions determine the installation vector, while chronological validation establishes its operational feasibility with cross-day energy continuity.

\section{Harmonic Propagation and Active ESS Support}\label{sec:harmonic}
\subsection{IBR harmonic sources and passive loads}
IBRs are the harmonic-current sources in the case study, while loads enter the harmonic network through their passive impedances. The emission ratio $\alpha_h$ relates the magnitude of each modeled harmonic current to the fundamental IBR current. At generator bus $j$, let $P_{jt}=\sum_{g\in\cG_j}p_{gt}$ and let $I^{0,\mathrm{ex}}_{jt}$ denote the exact fundamental injection magnitude. At unity fundamental power factor, they satisfy
\begin{equation}
P_{jt}=\sqrt{v_{jt}}I^{0,\mathrm{ex}}_{jt}.\label{noncovex}
\end{equation}
The adaptive DT construction in Section~\ref{sec:dt} provides a certified affine representation of this nonlinear voltage--current relationship for use in the mixed-integer conic model.

\subsection{Frequency-dependent network representation}
For each modeled order $h$ and hour $t$, an admittance matrix $Y^h_t$ is assembled from the active network branches, transformer ratios, charging shunts, and passive load impedances. A branch with harmonic parameters $R^h_\ell,X^h_\ell$ has
\begin{equation}
y^h_\ell=(R^h_\ell+\jmathu X^h_\ell)^{-1}.
\end{equation}
Here $\jmathu^2=-1$, and $R^h_\ell,X^h_\ell$ are the resistance and reactance evaluated at order $h$. Charging susceptances enter as bus-side shunts. Hourly passive load admittance is obtained from a nominal-voltage impedance $Z^1_{it}=1/(p^d_{it}-\jmathu q^d_{it})$ and the series model
\begin{equation}
Y^{h,\mathrm{load}}_{it}
=\left(\Re Z^1_{it}+\jmathu h\Im Z^1_{it}\right)^{-1}.\label{eq:loadimp}
\end{equation}
Zero-load buses contribute zero load admittance. The matrix $Z^h_t=(Y^h_t)^{-1}$ is computed before each optimization and enters the harmonic constraints as a given parameter. 

\subsection{Current emission and local cancellation}
The exact source current can be recovered from the operating point as
\begin{equation}
I^{0,\mathrm{ex}}_{jt}=\frac{P_{jt}}{\sqrt{v_{jt}}}.\label{eq:iexact}
\end{equation}
The optimization variable $I^0_{jt}$ bounds this current from above through the certified envelope. For the compensated order $h_c=5$, ESS unit $s$ supplies a nonnegative harmonic-current magnitude $k_{st}$. Local phase tracking is assumed to oppose the colocated IBR's fifth-harmonic current. The residual magnitude entering the network bound is
\begin{align}
J^h_{jt}&=\alpha_h I^0_{jt}
-\mathbf{1}_{h=5}\sum_{s\in\cS_j}k_{st},\label{eq:residual}\\
0\le\sum_{s\in\cS_j}k_{st}&\le\alpha_5\frac{P_{jt}}{\overline V}.\label{eq:nocancel}
\end{align}
The indicator $\mathbf{1}_{h=5}$ equals one for the fifth harmonic and zero otherwise. The upper bound in~\eqref{eq:nocancel} prevents over-cancellation even when $I^0_{jt}$ exceeds the exact current: $P_{jt}/\overline V\le I^{0,\mathrm{ex}}_{jt}\le I^0_{jt}$. Compensation is therefore restricted to buses containing a modeled source.

\subsection{Converter capability coupled to ESS dispatch}
Let $v_{st}$ denote the squared voltage at the bus hosting ESS $s$. At unity fundamental power factor, its actual fundamental current is $|p^e_{st}|/\sqrt{v_{st}}$. The following conservative current-rating constraint reserves shared converter capability for that current and the harmonic service:
\begin{equation}
\left\|\begin{bmatrix}p^e_{st}/\underline V\\ k_{st}\end{bmatrix}\right\|_2
\le\bar I_s.\label{eq:converter}
\end{equation}
Charging and discharging use part of the converter current rating, leaving the remaining capability available for harmonic compensation. Equation~\eqref{eq:soc} restricts when charging and discharging can occur, while~\eqref{eq:converter} limits the harmonic current available at each of those operating points.

\subsection{Network-wide harmonic limits}
A phase-independent magnitude bound represents network-wide harmonic propagation through the triangle inequality:
\begin{equation}
\bar V^h_{it}=\sum_{j\in\cG_{\!B}}|Z^h_{ij,t}|J^h_{jt},\label{eq:propagation}
\end{equation}
Here $\cG_{\!B}$ is the set of generator buses, and $Z^h_{ij,t}$ is the transfer-impedance entry relating current at bus $j$ to voltage at bus $i$. Local cancellation in~\eqref{eq:residual} is applied before taking the network-wide magnitude bound.

With per-unit distortion limits $L_h,L_{\mathrm{THD}}$ and a numerical margin $0<\gamma<1$, the constraints are
\begin{align}
(\bar V^h_{it})^2&\le(\gamma L_h)^2 v_{it},\quad h\in\cH,\label{eq:ihdcon}\\
\sum_{h\in\cH}(\bar V^h_{it})^2&\le(\gamma L_{\mathrm{THD}})^2v_{it}.\label{eq:thdcon}
\end{align}
These are convex quadratic/SOC-representable inequalities. For reporting, the exact injection current~\eqref{eq:iexact} replaces $I^0$ in~\eqref{eq:residual}--\eqref{eq:propagation}, giving $\bar V^{h,\mathrm{ex}}$. The plotted quantities are
\begin{align}
\mathrm{IHD}_{it}^h(\%)&=100\frac{\bar V^{h,\mathrm{ex}}_{it}}{\sqrt{v_{it}}},\label{eq:ihdreport}\\
\mathrm{THD}_{it}(\%)&=100\frac{\sqrt{\sum_{h\in\cH}(\bar V^{h,\mathrm{ex}}_{it})^2}}{\sqrt{v_{it}}}.\label{eq:thdreport}
\end{align}
The reported IHD and THD are network magnitude-bound indicators evaluated with recalculated source currents. THD aggregates the six modeled harmonic orders.

\section{Certified Adaptive DT Envelope}\label{sec:dt}
DT-based representations have been applied to power-system relationships, including the Q--V curve used for voltage-stability assessment~\cite{jia2024learning}. Here, a regression tree maps the input $(v,I)$ to power on the analytic surface $f(v,I)=\sqrt v I$. OPF operating points train the tree and determine its leaf partitions. After partitioning, a separate analytic grid within each leaf supplies the points for affine fitting and certification.

\subsection{OPF training data and complete leaf regions}
The training data comprise voltage, current and power values collected from OPF solutions under different load and generation conditions. The data are randomly divided into training, validation and test sets using a 70/15/15 split. Candidate regression trees have depths 3--8 and minimum leaf populations of 50, 100, 200, or 500. Among candidates with at most 16 leaves, validation mean absolute error (MAE) of the power prediction determines selection, with fewer leaves breaking ties. Tree selection and subsequent affine-envelope certification are separate steps.

Each root-to-leaf path defines a complete rectangle
$\mathcal R_r=[v_r^-,v_r^+]\times[I_r^-,I_r^+]$.
The rectangles cover $[\underline V^2,\overline V^2]\times[0,\bar I^0]$ with disjoint interiors; shared boundaries can be assigned to either adjacent leaf. The selected tree has 16 current-direction strips spanning the voltage range.

\subsection{Affine fitting and analytic certification}
After the tree partitions are fixed, a $3\times3$ tensor grid is generated within each leaf. The power value at each grid point is evaluated analytically as $P=\sqrt v I$. These nine points are used to fit an initial affine plane $a_r v+b_r I+c_r$ by least squares. Define its maximum residual
\begin{equation}
\rho_r=\max_{(v,I)\in\mathcal R_r}
\{a_rv+b_rI+c_r-\sqrt v I\}.\label{eq:resmax}
\end{equation}
This maximum can be evaluated at the four corners. Indeed, the residual is affine in $I$ for fixed $v$, so one of the two current endpoints attains its maximum. At either endpoint $I\ge0$, its second derivative in $v$ is $I/(4v^{3/2})\ge0$; a convex univariate function attains a maximum on an interval at an endpoint. Consequently, shifting the intercept yields
\begin{equation}
\underline f_r(v,I)=a_rv+b_rI+c_r-\max(0,\rho_r)-\varepsilon_c
\le\sqrt v I,\label{eq:certificate}
\end{equation}
Here $\varepsilon_c=10^{-12}$ is a downward margin, and the certified intercept is $\widehat c_r=c_r-\max(0,\rho_r)-\varepsilon_c$. Certification applies throughout the leaf, including points outside the fitting grid.

If the shifted fit has a nonpositive current coefficient or becomes negative at a corner, the implementation uses a chord-based affine underestimator. Let $x^-=\sqrt{v_r^-}$, $x^+=\sqrt{v_r^+}$, and define the chord of $\sqrt v$ on the leaf interval as
\begin{equation}
q_r(v)=x^-+\frac{x^+-x^-}{v_r^+-v_r^-}(v-v_r^-).
\end{equation}
The two affine lower-bound candidates are
\begin{align}
L_{r1}&=I_r^-q_r(v)+x^-I-I_r^-x^-,\\
L_{r2}&=I_r^+q_r(v)+x^+I-I_r^+x^+.
\end{align}
The candidate with the larger value at the region center is selected and its corner residual is checked. Its coefficients replace $a_r,b_r,\widehat c_r$ in the certified plane. Both candidates combine the chord lower bound on $\sqrt v$ with an affine lower bound on its product with nonnegative current.

For either construction, enforcing $P_{jt}\le\underline f_r(v_{jt},I^0_{jt})$ gives $I^0_{jt}\ge P_{jt}/\sqrt{v_{jt}}$. This ordering gives a conservative source-current magnitude. Combining this ordering with~\eqref{eq:nocancel} also preserves nonnegative conservative residuals when ESS cancellation is enabled.

This ordering is preserved when multiplying by $\alpha_h\ge0$, subtracting the same feasible compensation, and summing residual currents with nonnegative weights $|Z^h_{ij,t}|$. Hence replacing $I^0$ by the recalculated current can only reduce the harmonic-voltage bound at the same operating point.

\subsection{Disaggregated mixed-integer formulation}
For each generator bus and hour, introduce leaf selectors $z_{jtr}\in\{0,1\}$ and disaggregated variables $v_{jtr},I_{jtr}$. With the certified plane coefficients denoted $a_r,b_r,\widehat c_r$,
\begin{align}
\sum_r z_{jtr}&=1,\quad v_{jt}=\sum_r v_{jtr},\quad I^0_{jt}=\sum_r I_{jtr},\label{eq:hullsum}\\
v_r^-z_{jtr}&\le v_{jtr}\le v_r^+z_{jtr},\label{eq:hullv}\\
I_r^-z_{jtr}&\le I_{jtr}\le I_r^+z_{jtr},\label{eq:hulli}\\
P_{jt}&\le\sum_r(a_rv_{jtr}+b_rI_{jtr}+\widehat c_rz_{jtr}).\label{eq:hullpower}
\end{align}
Exactly one leaf is active for each bus and hour. Inactive leaves have zero disaggregated variables; the active leaf carries the full voltage and current and enforces its certified plane. Thus integral selectors represent the union of the certified leafwise sets using the leaf bounds directly. Combined with the network, ESS and OLTC constraints, it yields a MISOCP with continuous conic subproblems.

The current domain uses $\bar I^0=0.2$ p.u. with 16 leaf selectors per generator-bus-hour. Corner certification and independent analytic-point checks verify the lower-envelope construction. Integral leaf reconstruction checks verify its implementation at the evaluated operating points.

\section{Case study}
\subsection{System Description}
\begin{figure}[htbp]
\centering
\includegraphics[width=\columnwidth]{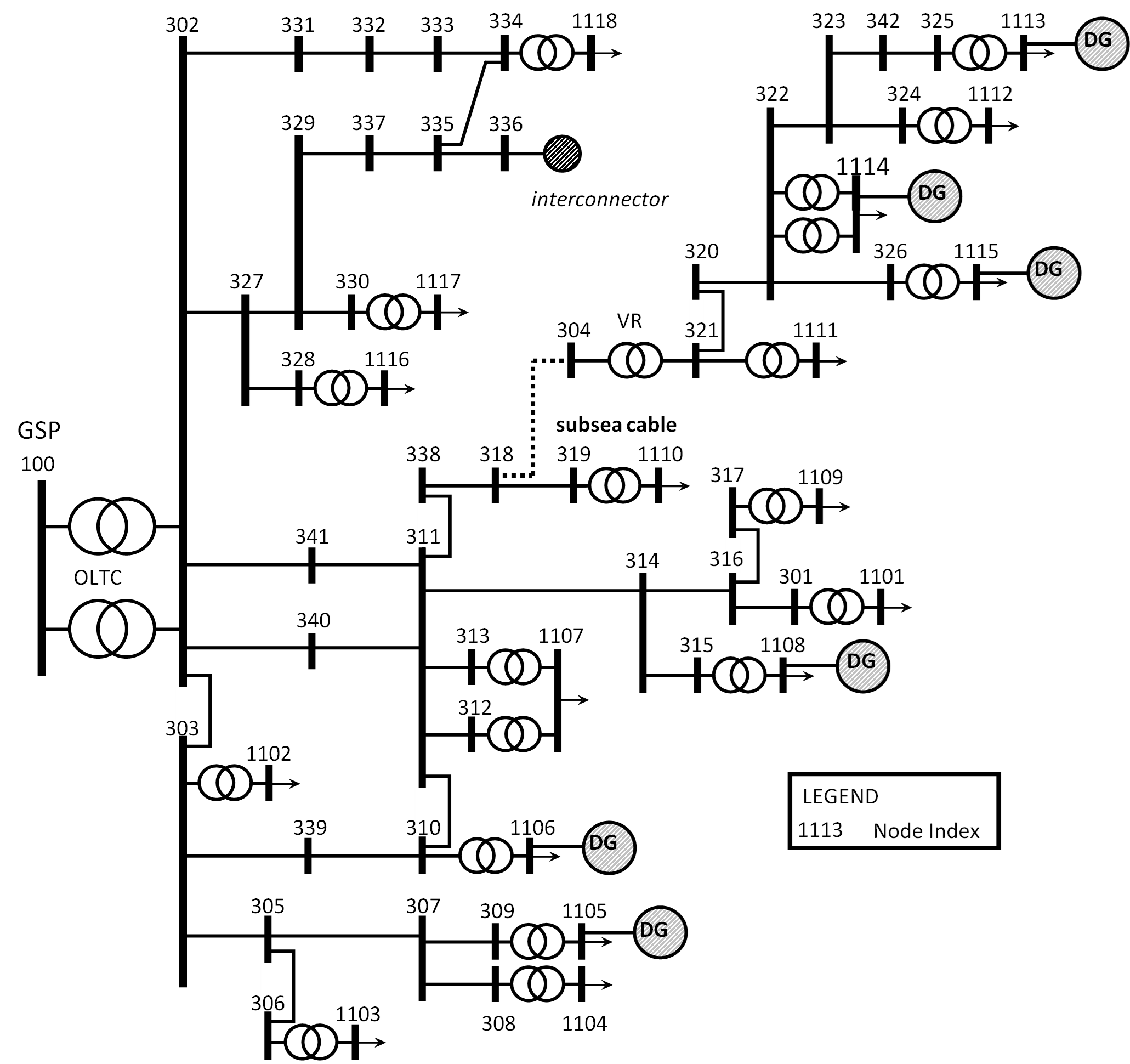}
\caption{U.K. GDS EHV1 Network \cite{ochoa2010minimizing}.}
\label{Network}
\end{figure}

The analysis is applied to the generic U.K. medium-voltage distribution network \cite{ochoa2010minimizing} shown in Fig. \ref{Network}. The feeders are supplied via a grid supply point by two identical 30-MVA, 132/33-kV transformers. The voltage limits are set at ±6\% of the nominal value, consistent with U.K. standards. The total peak demand of the system is 35 MW.

\begin{figure}[!t]
\centering
\includegraphics[width=\columnwidth]{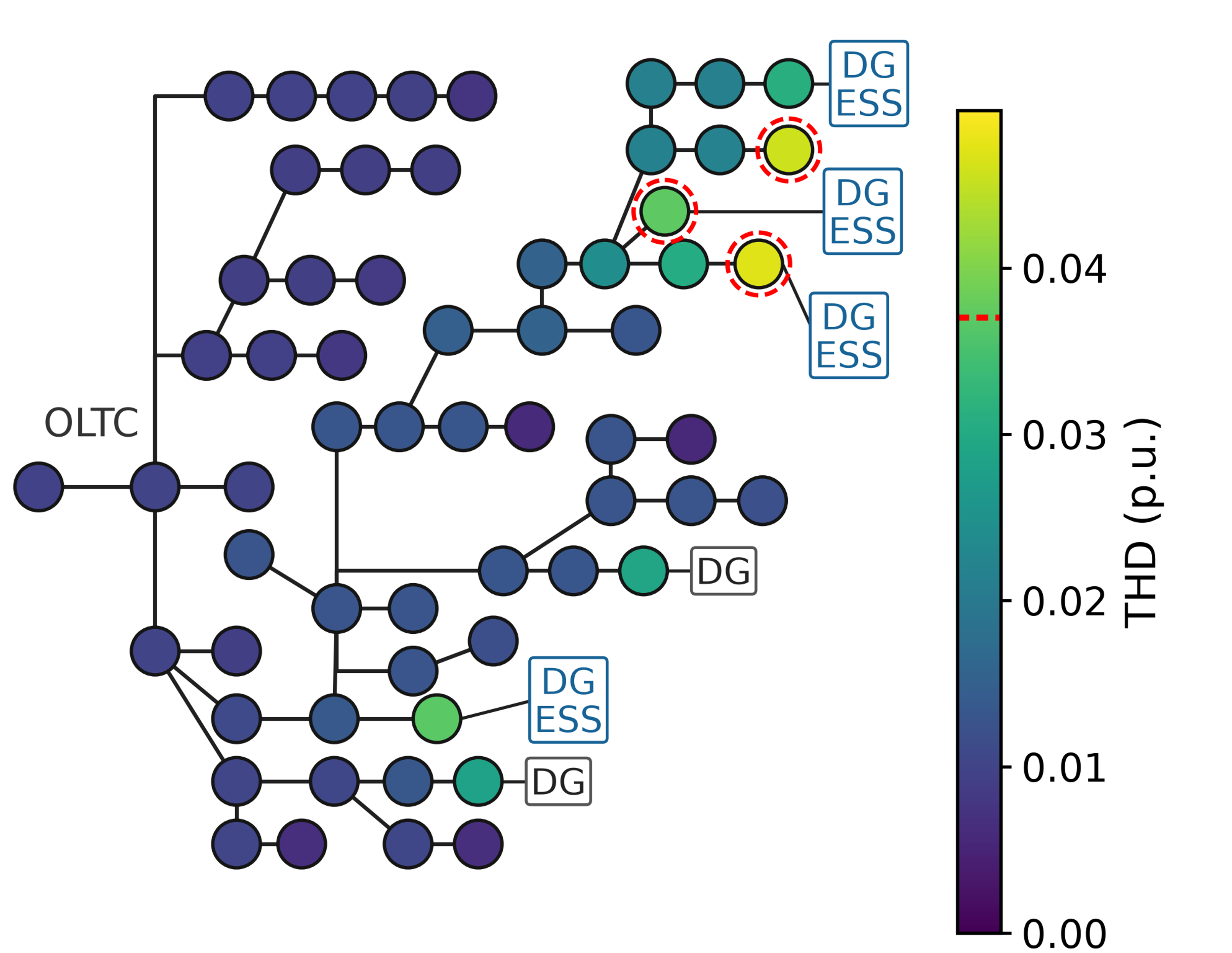}
\caption{Buswise maximum THD bounds for the OPF reference on the day of its annual maximum THD.}
\label{fig:OPFTHD}
\end{figure}

\begin{figure}[!t]
\centering
\includegraphics[width=\columnwidth]{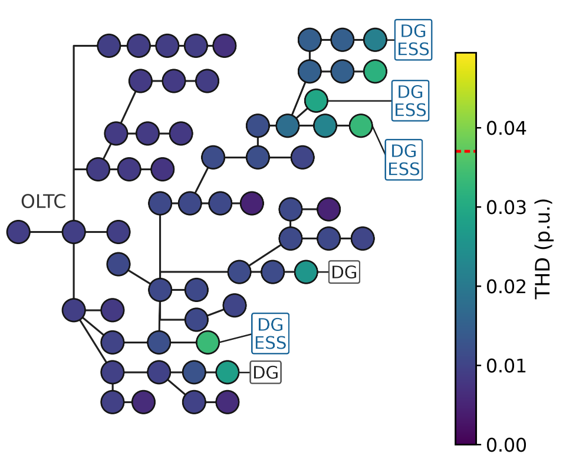}
\caption{Buswise maximum THD bounds for HOPF without support on the same day as the OPF reference map.}
\label{fig:HOPFTHD}
\end{figure}

\begin{figure}[!b]
    \centering
    \includegraphics[width=0.95\columnwidth]{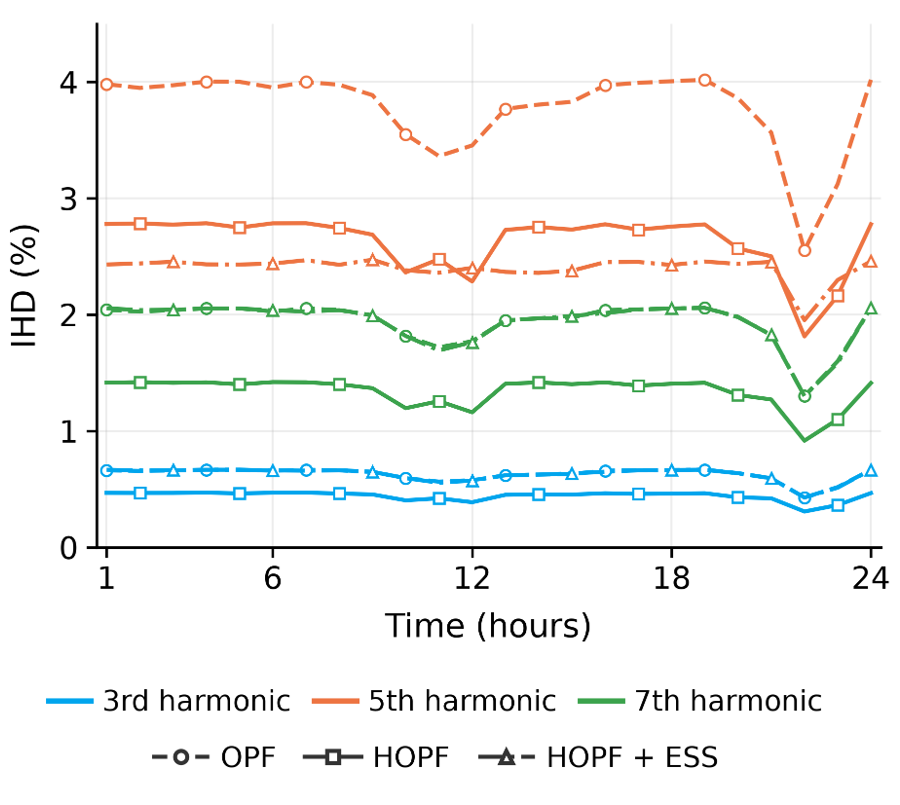}
    \caption{Third-, fifth- and seventh-order IHD bounds at Bus 1112 on the day of the annual OPF maximum THD. All three scenarios use the same date and their annually validated capacity configurations. Colours distinguish harmonic orders; line styles and markers distinguish scenarios.}
    \label{fig:IHD2}
\end{figure}

\begin{figure*}[!t]\centering
\includegraphics[width=.84\textwidth]{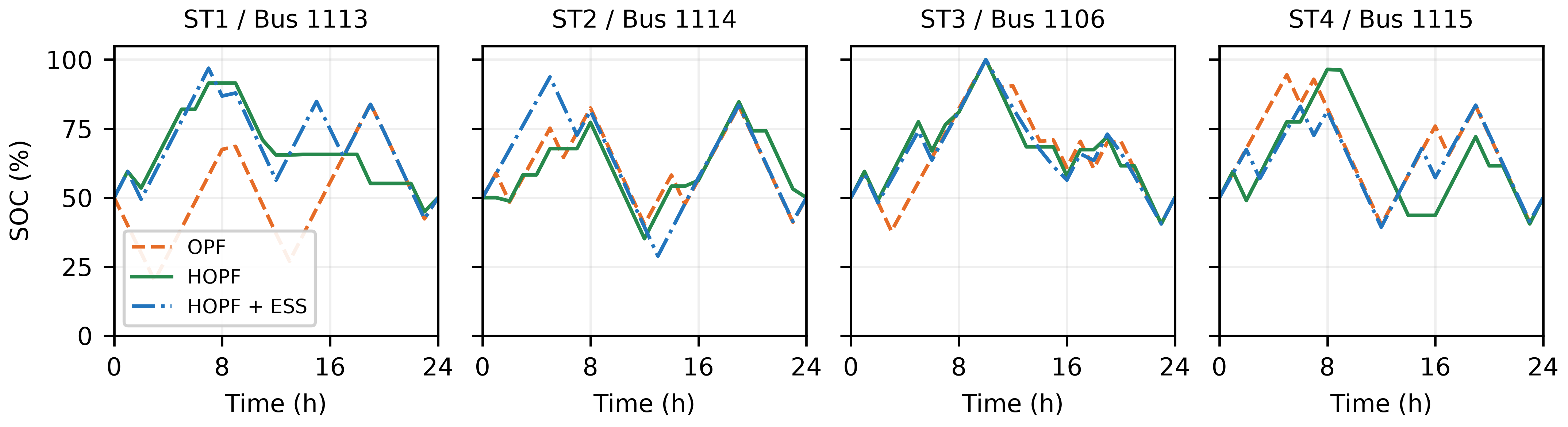}
\caption{SOC trajectories of the four ESS units for the three scenarios on the common display day selected from the annual OPF THD maximum.}
\label{fig:soc24}
\par\medskip
\includegraphics[width=.84\textwidth]{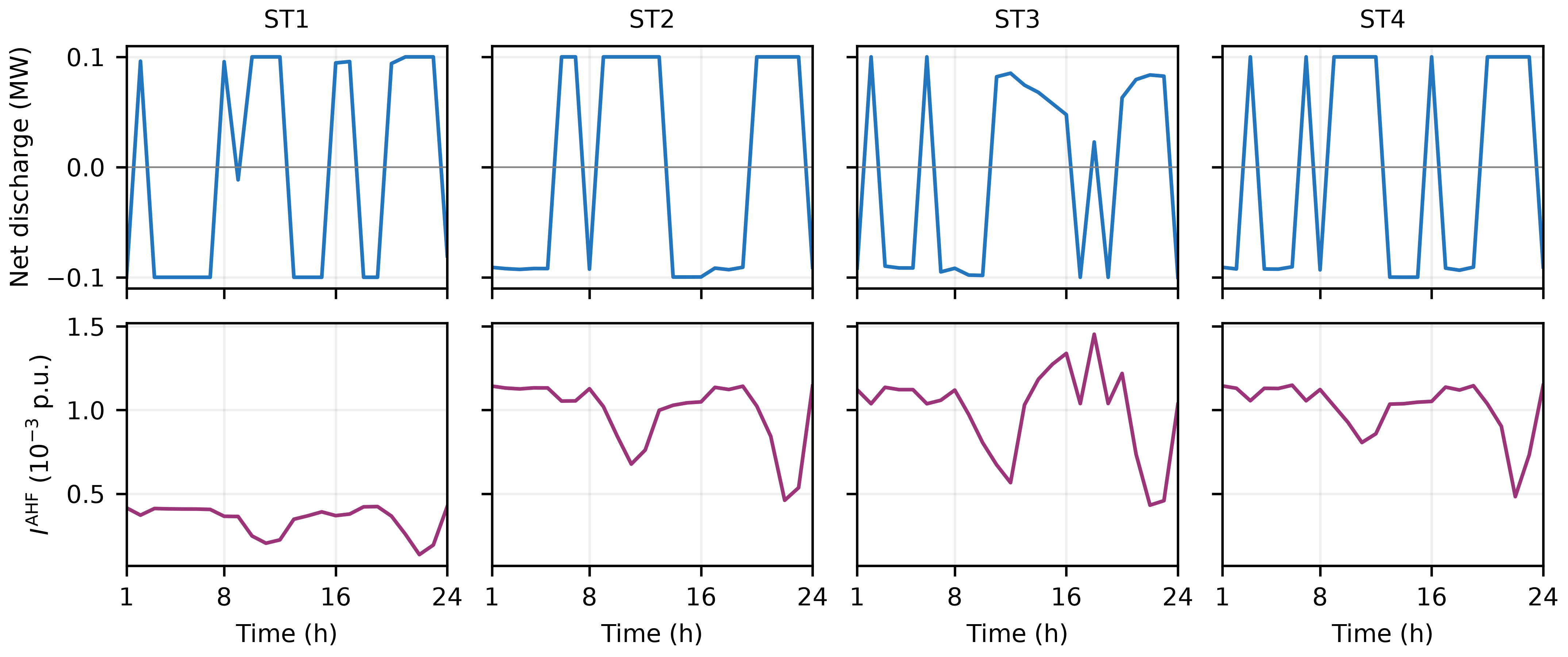}
\caption{Coordinated ESS operation under the recovered OPF capacity configuration on the common display day: net active power (top) and fifth-harmonic compensation current (bottom, $I^{\mathrm{AHF}}=k_{st}$). Positive active power indicates discharging.}
\label{fig:esscoupling}\end{figure*}

Figs.~\ref{fig:OPFTHD}--\ref{fig:HOPFTHD} compare buswise THD, and Fig.~\ref{fig:IHD2} compares hourly IHD at Bus 1112. Figs.~\ref{fig:soc24} and~\ref{fig:esscoupling} show the corresponding SOC trajectories and coordinated ESS schedules.

\setlength{\dbltextfloatsep}{12pt plus 2pt minus 2pt}
\setlength{\dblfloatsep}{8pt plus 2pt minus 2pt}

Six distributed wind farms are connected to the network. These are divided into two groups with each group using a different wind profile. A full year of hourly profiles is processed using the daily optimization and chronological annual-validation workflow in Section~\ref{sec:annual}. Each daily capacity subproblem is solved to optimality; the aggregated capacity vector is then validated over the full year. The model is built in Julia and solved by Gurobi 11.0.1 \cite{gurobi} on a Windows 11 computer with an Intel i7-12700 CPU and 32 GB RAM. ESS parameters are listed in Table~\ref{tab:ess}. All four ESS units are present in every scenario; only their harmonic support function is disabled in OPF and uncompensated HOPF.

\begin{table}[t]\centering
\caption{ESS parameters common to the three scenarios}\label{tab:ess}
\footnotesize
\begin{tabular}{ll}\toprule
Quantity & Value \\\midrule
ST1 / ST2 locations & Bus 1113 / Bus 1114 \\
ST3 / ST4 locations & Bus 1106 / Bus 1115 \\
Charge/discharge power, each & 0.1 MW \\
Energy capacity, each & 1 MWh \\
Converter current rating, each & 0.0015 p.u. \\
Nominal converter rating, each & 0.15 MVA at 1 p.u. voltage \\
Charge/discharge efficiency & 0.95 / 0.95 \\
Daily capacity subproblem SOC & 50\% / 50\% \\
Compensated harmonic order & Fifth only \\\bottomrule
\end{tabular}\end{table}

Table~\ref{tab:spectrum} lists the assumed normalized harmonic
current spectrum of the IBRs and the IHD limits adopted in this study. The limit values are inspired by the 33-kV planning levels in ENA EREC G5 Issue 5~\cite{ena2020g5}. The THD limit is 3.7\%.

\begin{table}[t]\centering
\caption{Harmonic spectrum and assumed IHD limits}\label{tab:spectrum}
\footnotesize
\begin{tabular}{lrrrrrr}\toprule
Order $h$ & 3 & 5 & 7 & 11 & 13 & 17 \\\midrule
$100\alpha_h$ (\%) & 0.636 & 2.358 & 1.104 & 0.180 & 0.156 & 0.024 \\
$100L_h$ (\%) & 2.6 & 2.8 & 2.8 & 1.9 & 1.8 & 1.4 \\\bottomrule
\end{tabular}\end{table}

The proposed method and its two evaluation references use the same ESS equipment and optimize their own OLTC settings:
\begin{enumerate}
\item \emph{OPF reference}: maximize capacity under fundamental network and ESS energy constraints with $k_{st}=0$. Harmonic indicators are evaluated from the resulting operating points.
\item \emph{HOPF without support}: maximize capacity with the DT envelopes and harmonic limits, retaining $k_{st}=0$. The capacity search is capped at the OPF reference total.
\item \emph{HOPF + active ESS}: jointly schedule charging, discharging and fifth-harmonic compensation while maximizing total capacity. Each generator capacity has an upper bound equal to its corresponding OPF reference value. Reaching those bounds establishes recovery of the reference allocation.
\end{enumerate}
The reported capacities are the common annual configurations obtained from the componentwise minima of the daily optimal solutions and verified through chronological full-year operation.

\subsection{Capacity restriction and recovery}
Table~\ref{tab:results} reports the annually validated capacity of each scenario and its full-year operating extrema. Harmonic indicators use the recalculated source currents defined in~\eqref{eq:iexact}. The OPF reference accommodates \OPFCap{} MW, with IHD and THD bounds exceeding the adopted thresholds. HOPF without support yields an annually validated capacity of \HOPFCap{} MW, \CapReduction\% below the reference. Its maximum IHD is close to the corresponding limit, while THD retains margin. 

\begin{table}[t]\centering
\caption{Annually validated capacities and full-year operating extrema}\label{tab:results}\footnotesize
\setlength{\tabcolsep}{3pt}
\begin{tabular}{lrrr}\toprule
Quantity & OPF & HOPF & HOPF + ESS \\\midrule
Annual capacity (MW) & \OPFCap & \HOPFCap & \OPFCap \\
Maximum IHD (\%) & \OPFIHD & \HOPFIHD & \AHFIHD \\
Maximum THD (\%) & \OPFTHD & \HOPFTHD & \AHFTHD \\
Minimum voltage (p.u.) & \OPFVmin & \HOPFVmin & \AHFVmin \\
Maximum voltage (p.u.) & \OPFVmax & \HOPFVmax & \AHFVmax \\
\bottomrule
\end{tabular}
\end{table}

With active ESS compensation, every generator reaches its OPF capacity upper bound and the complete reference allocation is recovered. The full-year maximum IHD and THD bounds are \AHFIHD\% and \AHFTHD\%, respectively. Chronological validation establishes feasibility of this common capacity configuration over the year. The resulting capacity configurations are also validated using nonlinear AC power flow. 

Storage energy scheduling is available in both references, so the recovered capacity is attributable to enabling compensation within the same storage ratings. Reaching every per-generator upper bound also preserves the spatial allocation of the OPF reference.

\subsection{Spatial and hourly harmonic behavior}

The displayed day is selected as the day containing the maximum THD over all buses and hours of the annual OPF reference. All three scenarios use this same date for the hourly comparisons. Figs.~\ref{fig:OPFTHD}--\ref{fig:HOPFTHD} show each bus's daily maximum THD on that date, using a common color scale. Red rings identify buses exceeding the assumed threshold, and the red dashed color-bar marker denotes 3.7\%. On the displayed day, the OPF THD maximum occurs at Bus 1115 in hour 7, while the largest IHD occurs at the fifth harmonic at Bus 1112 in hour 24. Bus 1112 carries constant active and reactive power demand and has no colocated IBR.

The different locations and hours follow from the frequency-dependent transfer impedances and the aggregation of harmonic orders in THD. Equation~\eqref{eq:propagation} sums the contributions of all modeled source buses at each observation bus. Local compensation therefore changes the distortion bound elsewhere in the network, including at load buses such as Bus 1112.

IHD constrains each harmonic separately, whereas THD constrains their root-sum-square magnitude. All six modeled orders remain in these constraints when fifth-harmonic compensation is enabled. The maps express THD in per unit; their 0.037 threshold corresponds to the 3.7\% annual-results limit.

Fig.~\ref{fig:IHD2} compares third-, fifth-, and seventh-order IHD at Bus 1112 across all three scenarios. The fifth harmonic is the principal local restriction on this day. The curves compare the hourly distortion produced by each scenario's common annual capacity and optimized dispatch. The compensated case operates with the recovered OPF allocation, so its harmonic performance is assessed at the higher generation capacity.

\subsection{ESS scheduling and compensation capability}
Fig.~\ref{fig:soc24} compares the SOC trajectories of the four ESS units on the common display day. Storage remains available for energy scheduling in all three cases. The different trajectories reflect each capacity configuration and its operating constraints; in the compensated case, dispatch also determines the converter current left for harmonic support.

Fig.~\ref{fig:esscoupling} shows the coordinated active-power and fifth-harmonic compensation schedules under the recovered OPF capacity configuration. Positive active power denotes discharging. The converter constraint links both trajectories: larger active power consumes more of the available current rating. The maximum converter-current utilization in the reported solution is approximately 99.92\%, showing that the combined services make use of nearly the full converter rating. The four units' total active-power rating, energy capacity, and nominal converter rating are 0.4 MW, 4 MWh, and 0.6 MVA, respectively. The capacity recovery is therefore obtained by coordinating the two services within this fixed storage configuration.

Available compensation is limited by both converter headroom and the colocated source current. Low active-power exchange increases headroom, while~\eqref{eq:nocancel} caps compensation at the local fifth-harmonic emission. Joint scheduling respects both restrictions alongside the storage energy trajectory.

\FloatBarrier
\section{Conclusion}
The proposed HOPF with active ESS support combines certified DT current envelopes, intertemporal storage dispatch and network-wide harmonic limits in a MISOCP. Sharing the converter-current rating between active power and compensation links the two ESS services.

Daily optimal capacities are aggregated componentwise and validated through chronological full-year operation. In the case study, HOPF without support yields \HOPFCap{} MW, compared with the \OPFCap{} MW OPF reference. With active ESS support, every generator reaches its OPF capacity upper bound while the adopted harmonic bounds are met. Nonlinear AC power-flow validation confirms capacity feasibility. Coordinating the two ESS services therefore recovers the reference allocation within the existing storage and converter ratings.

\bibliographystyle{IEEEtran}
\bibliography{bibtex/bib/IEEEexample}

@manual{ena2020g5,
  title        = {Engineering Recommendation {G5}, Issue 5:
                  Harmonic Voltage Distortion and the Connection
                  of Harmonic Sources and/or Resonant Plant to
                  Transmission Systems and Distribution Networks
                  in the {United Kingdom}},
  organization = {{Energy Networks Association}},
  address      = {London, U.K.},
  year         = {2020},
  month        = jun,
  url          = {https://dcode.org.uk/assets/uploads/ENA_EREC_G5_Issue_5__2020__2.pdf}
}

@article{low2014convex,
  title={Convex relaxation of optimal power flow—Part I: Formulations and equivalence},
  author={Low, Steven H},
  journal={IEEE Transactions on Control of Network Systems},
  volume={1},
  number={1},
  pages={15--27},
  year={2014},
  publisher={IEEE}
}

@article{misra2017harmonic,
  title={Harmonic distortion minimization in power grids with wind and electric vehicles},
  author={Misra, Ritam and Paudyal, Sumit and Ceylan, O{\u{g}}uzhan and Mandal, Paras},
  journal={Energies},
  volume={10},
  number={7},
  pages={932},
  year={2017},
  publisher={Mdpi}
}

@article{yang2023opendss,
  title={{OpenDSS}-based harmonic power flow analysis for distribution systems with passive power filters},
  author={Yang, Nien-Che and Hsu, Yu-Wei},
  journal={IEEE Access},
  volume={11},
  pages={69190--69203},
  year={2023},
  publisher={IEEE}
}

@article{sun2022distribution,
  title={Distribution network hosting capacity assessment: Incorporating probabilistic harmonic distortion limits using chance constrained optimal power flow},
  author={Sun, Wei and Harrison, Gareth P},
  journal={IET Smart Grid},
  volume={5},
  number={2},
  pages={63--75},
  year={2022},
  publisher={Wiley Online Library}
}

@misc{gurobi,
  author = {{Gurobi Optimization, LLC}},
  title = {{Gurobi Optimizer Reference Manual}},
  year = 2024,
  url = "https://www.gurobi.com"
}

@article{wang2018optimal,
  title={Optimal operation of active distribution network involving the unbalance and harmonic compensation of converter},
  author={Wang, Jian and Zhou, Niancheng and Ran, Yunhao and Wang, Qianggang},
  journal={IEEE Transactions on Smart Grid},
  volume={10},
  number={5},
  pages={5360--5373},
  year={2019},
  doi={10.1109/TSG.2018.2880680},
  publisher={IEEE}
}

@article{xiao2022optimal,
  title={Optimal power quality compensation of energy storage system in distribution networks based on unified multi-phase {OPF} model},
  author={Xiao, Xianyong and Li, Zhuocheng and Wang, Yang and Zhou, Yaxiang and Liu, Ke},
  journal={IEEE Transactions on Smart Grid},
  volume={13},
  number={3},
  pages={1873--1887},
  year={2022},
  publisher={IEEE}
}

@article{geth2022harmonic,
  title={Harmonic optimal power flow with transformer excitation},
  author={Geth, Frederik and Van Acker, Tom},
  journal={Electric Power Systems Research},
  volume={213},
  pages={108604},
  year={2022},
  publisher={Elsevier}
}

@inproceedings{xue2023resonance,
  title={Resonance Participation Factor Evaluation for Renewable Energy Systems Based on Impedance Scanning},
  author={Xue, Yinglin and Feng, Ning and Wang, Yue},
  booktitle={2023 IEEE 7th Conference on Energy Internet and Energy System Integration (EI2)},
  pages={1858--1863},
  year={2023},
  organization={IEEE}
}

@article{ornelas2019optimal,
  title={Optimal dynamic harmonic extraction and suppression in power conditioning applications},
  author={Ornelas-Tellez, Fernando and Rico-Melgoza, J Jesus and Morfin-Maga{\~n}a, Rodrigo and Ramos-Paz, Serafin},
  journal={IEEE Transactions on Industrial Electronics},
  volume={67},
  number={9},
  pages={7909--7918},
  year={2020},
  doi={10.1109/TIE.2019.2944059},
  publisher={IEEE}
}

@article{pandi2012determining,
  title={Determining optimal location and size of distributed generation resources considering harmonic and protection coordination limits},
  author={Pandi, V Ravikumar and Zeineldin, H H and Xiao, Weidong},
  journal={IEEE Transactions on Power Systems},
  volume={28},
  number={2},
  pages={1245--1254},
  year={2013},
  doi={10.1109/TPWRS.2012.2209687},
  publisher={IEEE}
}

@article{taghvaie2023comprehensive,
  title={A comprehensive review of harmonic issues and estimation techniques in power system networks based on traditional and artificial intelligence/machine learning},
  author={Taghvaie, Amir and Warnakulasuriya, T and Kumar, Dinesh and Zare, Firuz and Sharma, Rahul and Vilathgamuwa, D Mahinda},
  journal={IEEE Access},
  volume={11},
  pages={31417--31442},
  year={2023},
  publisher={IEEE}
}

@article{jia2024learning,
  title={Learning Multiple Convex Voltage Stability Constraints for Unit Commitment},
  author={Jia, Hongyang and Hou, Qingchun and Yong, Pei and Teng, Fei and Strbac, Goran and Fang, Chen and Zhang, Ning},
  journal={IEEE Transactions on Power Systems},
  volume={40},
  number={1},
  pages={125--137},
  year={2025},
  doi={10.1109/TPWRS.2024.3402081},
  note={doi: 10.1109/TPWRS.2024.3402081},
  publisher={IEEE}
}

@article{singh2010grid,
  title={Grid interconnection of renewable energy sources at the distribution level with power-quality improvement features},
  author={Singh, Mukhtiar and Khadkikar, Vinod and Chandra, Ambrish and Varma, Rajiv K},
  journal={IEEE Transactions on Power Delivery},
  volume={26},
  number={1},
  pages={307--315},
  year={2011},
  doi={10.1109/TPWRD.2010.2081384},
  publisher={IEEE}
}

@article{singh2009power,
  title={Power system harmonics research: a survey},
  author={Singh, Girish Kumar},
  journal={European Transactions on Electrical Power},
  volume={19},
  number={2},
  pages={151--172},
  year={2009},
  publisher={Wiley Online Library}
}

@article{ochoa2010minimizing,
  title={Minimizing energy losses: Optimal accommodation and smart operation of renewable distributed generation},
  author={Ochoa, Luis F and Harrison, Gareth P},
  journal={IEEE Transactions on Power Systems},
  volume={26},
  number={1},
  pages={198--205},
  year={2011},
  doi={10.1109/TPWRS.2010.2049036},
  publisher={IEEE}
}

@article{wang2024pvess,
  author={Wang, Ying and Wang, Xinru and Li, Shunyi and Ma, Xiaoyang and Chen, Yunzhu and Liu, Shuming},
  title={Optimization model for harmonic mitigation based on {PV--ESS} collaboration in small distribution systems},
  journal={Applied Energy}, volume={356}, pages={122410}, year={2024},
  doi={10.1016/j.apenergy.2023.122410}
}

@article{antic2026envelopes,
  author={Anti{\'c}, Tomislav and Geth, Frederik and Keane, Andrew and Capuder, Tomislav},
  title={Calculation of Harmonic-Constrained Low Voltage Dynamic Operating Envelopes},
  journal={IEEE Transactions on Smart Grid}, volume={17}, number={4}, pages={3235--3247}, year={2026},
  doi={10.1109/TSG.2026.3660479}
}
\end{document}